\documentclass[runningheads]{llncs}
\usepackage[T1]{fontenc}
\usepackage{graphicx}
\usepackage{enumerate}
\usepackage{amsmath}

\usepackage{multirow}

\usepackage{array}
\usepackage{arydshln}

\usepackage{hyperref}
\usepackage{color}

\usepackage[misc]{ifsym}

\begin{document}
%
\title{CoMeta: Enhancing Meta Embeddings with Collaborative Information in Cold-start Problem of Recommendation}
\titlerunning{CoMeta}
%

\author{
Haonan Hu \and
Dazhong Rong\and
Jianhai Chen\textsuperscript{(\Letter)}\and
Qinming He\and
Zhenguang Liu}

%
\authorrunning{H. Hu et al.}
%
\institute{
Zhejiang University, Hangzhou, China\\
\email{\{huhaonan, rdz98, chenjh919, hqm\}@zju.edu.cn}, \email{\{liuzhenguang2008\}@gmail.com}
}
\maketitle              
\begin{abstract}
    The cold-start problem is quite challenging for existing recommendation models.
    Specifically, for the new items with only a few interactions, their ID embeddings are trained inadequately, leading to poor recommendation performance.
    Some recent studies introduce meta learning to solve the cold-start problem by generating meta embeddings for new items as their initial ID embeddings.
    However, we argue that the capability of these methods is limited, because they mainly utilize item attribute features which only contain little information, but ignore the useful collaborative information contained in the ID embeddings of users and old items.
    To tackle this issue, we propose CoMeta to enhance the meta embeddings with the collaborative information.
    CoMeta consists of two submodules: B-EG and S-EG.
    Specifically, for a new item: B-EG calculates the similarity-based weighted sum of the ID embeddings of old items as its base embedding; S-EG generates its shift embedding not only with its attribute features but also with the average ID embedding of the users who interacted with it.
    The final meta embedding is obtained by adding up the base embedding and the shift embedding.
    We conduct extensive experiments on two public datasets.
    The experimental results demonstrate both the effectiveness and the compatibility of CoMeta.

\keywords{Recommendation system \and Item cold-start problem \and Meta-learning \and Item ID embedding \and Deep learning}
\end{abstract}

\section{Introduction}
Fueled by the boom of internet, every day a large amount of information is produced.
Nowadays, information overload has become a severe problem affecting people’s daily life.
Recommendation systems, which filter useful information for users, can effectively relieve the information overload problem.
Therefore, recommendation systems have been deployed in various web services (\textit{e.g.}, short videos~\cite{dai2021poso}, e-commerce~\cite{zhao2022joint} and news portals~\cite{zhang2021unbert}).
Collaborative filtering~\cite{rendle2012bpr}, which learns users’ potential interests based on historical interactions, is one of the most widely used traditional recommendation methods.
Beyond collaborative filtering, many recent studies (\textit{e.g.}, Wide\&Deep~\cite{cheng2016wide} and DeepFM~\cite{guo2017deepfm}) utilize user’s features and items’ features additionally and apply deep neural networks (DNNs) to achieve better recommendation performance.
These deep recommendation models typically consist of two parts: an embedding layer and a multi-layer perceptron (MLP).
The embedding layer transforms the raw features into dense vectors, and the MLP captures more complicated semantics.

Despite the remarkable success of these models, they still suffer from the cold-start problem~\cite{schein2002methods}.
In these models, sufficient interaction data are required to train a reasonable ID embedding for each item.
However, in real scenarios many new items only have limited number of interactions, and hence their ID embeddings are insufficiently trained, which leads to a decrease in recommendation accuracy.
The cold-start problem is one of the major challenges for recommendation systems.

Lately, some methods have been proposed to address the cold-start problem.
DropoutNet~\cite{volkovs2017dropoutnet} and MTPR~\cite{du2020learn} dropout ID embeddings to improve the robustness of the recommendation model.
MetaEmb~\cite{pan2019warm} trains an embedding generator to generate a good initial ID embedding for each new item.
GME~\cite{ouyang2021learning} aims to build an item graph and learn a desirable initial ID embedding for each new item by utilizing its attribute features and other related old items' attribute features.
MWUF~\cite{zhu2021learning} proposes two meta networks to transform cold item ID embeddings into warm feature space.

The ID embeddings of users and old items are well-trained with sufficient interactions, and hence contain highly useful collaborative information.
However, all of the above methods do not take into account the collaborative information.
We argue the capability of these methods is still limited.
In this paper, we propose CoMeta to enhance the meta embeddings of new items with the collaborative information.
CoMeta consists of two components: Base Embedding Generator (B-EG) and Shift Embedding Generator (S-EG).
Under the common research settings, the cold-start problem is divided into two phases: cold-start phase (new items do not have any interactions in the training dataset) and warm-up phase (new items have a few interactions in the training dataset).
The calculation of both the base embeddings in B-EG and the average interacted user ID embeddings in S-EG relies on the interactions of new items.
Hence, our proposed CoMeta mainly aims to learn desirable initial ID embeddings for new items in the warm-up phase.
Moreover, since CoMeta is an extra module that generates initial ID embeddings for new items and has no influence on old items, CoMeta can be applied upon various recommendation models.

We summarize the main contributions of this paper as follows:
\begin{itemize}
    \item We propose a meta learning based method named CoMeta to solve the item cold-start problem by learning good initial ID embeddings for new items.
    CoMeta can be easily applied upon various recommendation models.
    \item In CoMeta, we propose two submodules, which are termed as B-EG and S-EG, to enhance the generated meta embeddings by utilizing the collaborative information.
    More Specifically, B-EG utilizes the collaborative information through the ID embeddings of old items.
    S-EG utilizes the collaborative information through the ID embeddings of the interacted users.
    \item We conduct extensive experiments on two large-scale public datasets.
    The experimental results show that CoMeta outperforms existing cold-start methods and can significantly improve the recommendation performance in addressing the cold-start problem.
\end{itemize}
\section{Related Work}

Recommendation systems are designed to recommend personalized items to users.
In early works, many researchers attempt to use traditional machine learning methods to make recommendations.
FM~\cite{rendle2010factorization}, which models low-order feature interactions, is one of the most widely used traditional machine learning recommendation methods. 
For its simplicity and effectiveness, many methods based on it are proposed,
However, the above methods can not capture high-order interactions. 
Recently, various deep learning models are proposed to solve this issue.
Wide\&Deep~\cite{cheng2016wide}, DeepFM~\cite{guo2017deepfm}, PNN~\cite{qu2016product} and DCN~\cite{wang2017deep} automatically cross high-order features to learn more information. 
DIN~\cite{zhou2018deep} and DIEN~\cite{zhou2019deep} utilize attention mechanism to capture users' interest based on users' historical interactions.
It is worth mentioning that some works also focus on the recommendation security~\cite{DBLP:conf/ijcai/RongHC22,rong2022fedrecattack}.

Despite the remarkable success of these models, they still suffer from the cold-start problem.
It is challenging to make reasonable recommendations for new users or new items with limited interaction data in recommendation systems.
To address the cold-start problem, some methods aim to learn a more robust model. 
DropoutNet~\cite{volkovs2017dropoutnet} and MTPR~\cite{du2020learn} dropout ID embeddings to reduce the model's dependency on them.
CC-CC~\cite{shi2019adaptive} replaces ID embeddings with random vectors.
Another way to solve this problem is to use variational autoencoder~\cite{kingma2013auto}.
CVAR~\cite{zhao2022improving} uses variational autoencoder to model the relationship between item ID and item features in the latent space.

Besides, many recent works introduce meta learning to tackle the cold-start problem.
Meta learning, also known as learning to learn, enables models to quickly adapt to new tasks with only few samples by utilizing prior knowledge from several related tasks.
MeLU~\cite{lee2019melu} generates the initial parameter of the model for all users based on Model-Agnostic Meta-Learning (MAML)~\cite{finn2017model} framework.
PAML~\cite{wang2021preference} introduces the social network to MeLU.
MetaEmb~\cite{pan2019warm} generates good initial ID embeddings for new items with their attribute features.
GME~\cite{ouyang2021learning} learns how to generate desirable initial ID embeddings for new items by considering both the new item attribute features and other related old items' attribute features.
MWUF~\cite{zhu2021learning} aims to train meta scaling and meta shifting networks to warm up cold ID embeddings.
\begin{figure}
    \centering
	\includegraphics[width=0.9\linewidth]{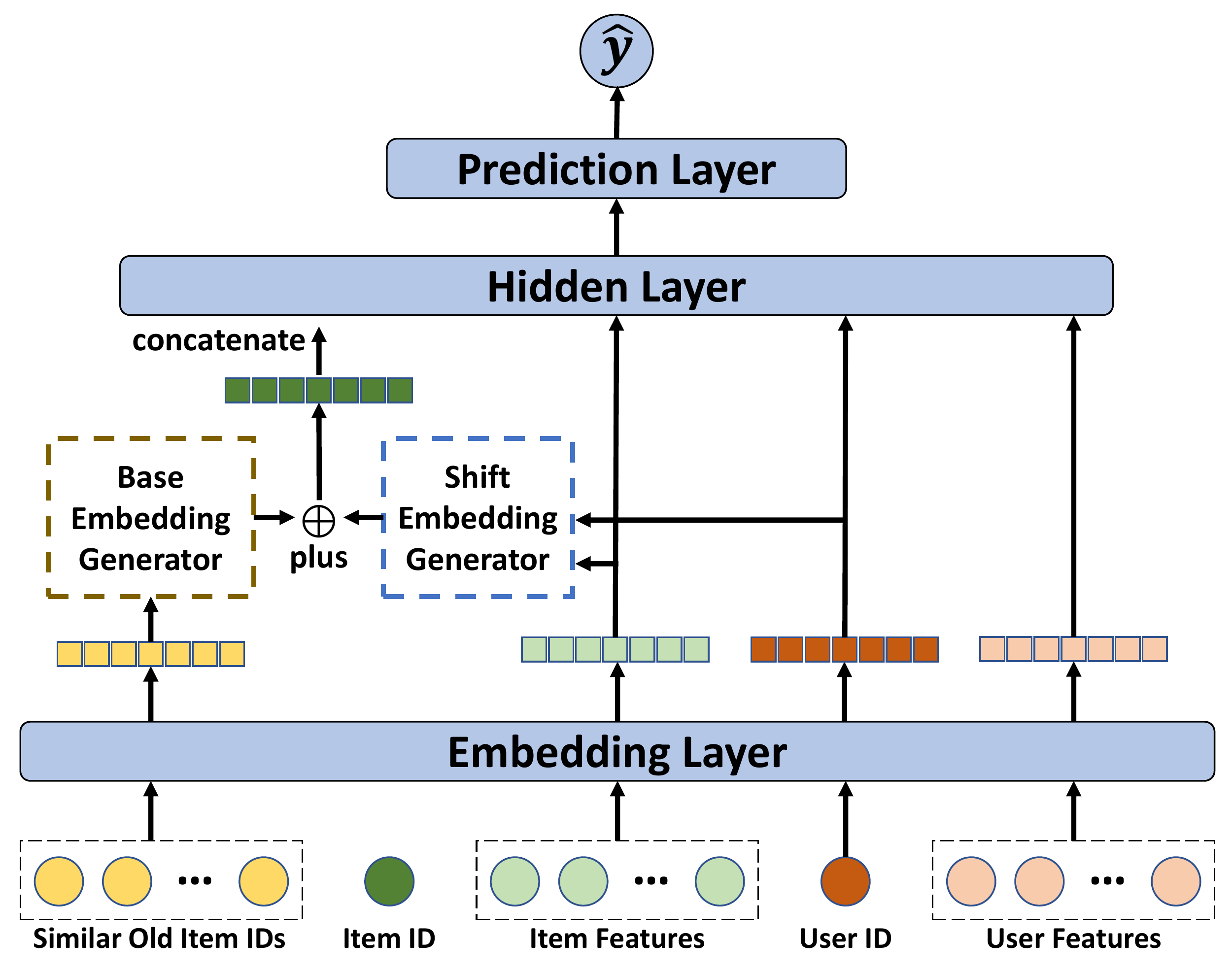} 
	\caption{The framework of CoMeta.}
	\label{fig:2} 
\end{figure}

\section{Method}
In this section, we propose CoMeta to generate desirable meta embeddings with collaborative information for new item IDs.

\subsection{Overview}
In this paper, only the new item ID embeddings are generated by the proposed CoMeta, and the old item ID embeddings can be used directly.
As shown in Figure \ref{fig:2}, our proposed CoMeta includes two components:
\begin{enumerate}[1)]
    \item Base Embedding Generator (B-EG).
    In B-EG, for each new item, we first compute the similarity scores between it and old items.
    Then we calculate the weighted sum of the ID embeddings of old items according to the similarity scores, and take the result as the generated base embedding.
    \item Shift Embedding Generator (S-EG).
    In S-EG, for each new item, we first calculate the average ID embeddings of the users who have interactions with it.
    And then we use deep neural networks to generate its shift embedding with its attribute features and the average interacted user ID embedding as the input.
\end{enumerate}
The final generated new item ID embedding can be denoted as:
\begin{equation}
    \mathbf{v}_{new} = \mathbf{v}_{BEG} + \mathbf{v}_{SEG},
\end{equation}
where $\mathbf{v}_{BEG}$ is the base embedding, and $\mathbf{v}_{SEG}$ is the shift embedding.

\begin{figure}
    \centering
	\includegraphics[width=0.6\linewidth]{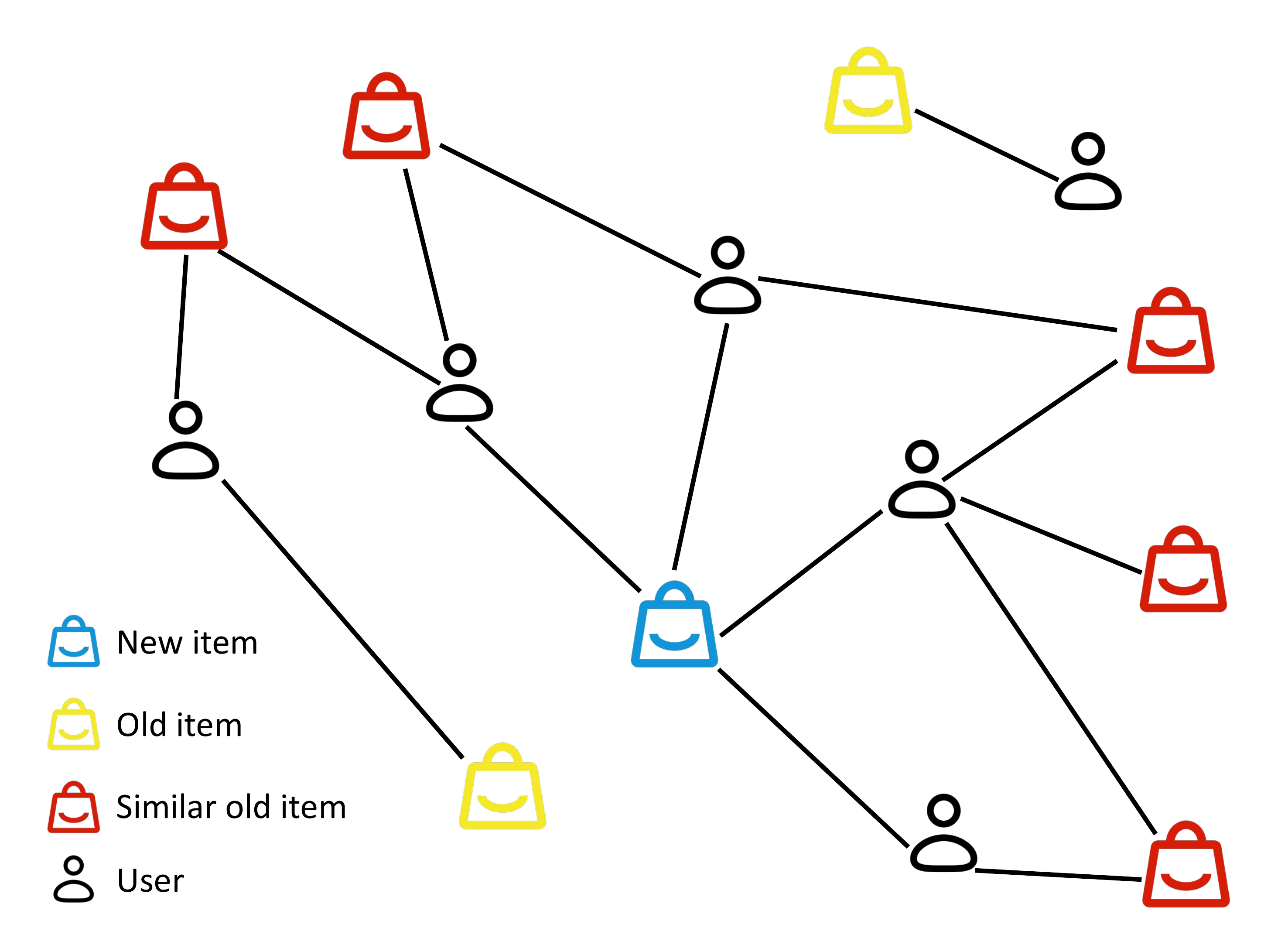} 
	\caption{Illustration of finding similar old items for a new item.}
	\label{fig:3} 
\end{figure}

\subsection{B-EG: Base Embedding Generator}
As the ID embeddings of new items are learned with only a few interactions, they lack the collaborative information to fit the recommendation model well.
However, the old item ID embeddings, which are trained with sufficient interactions, contain useful collaborative information.
Thus, we propose B-EG to capture the collaborative information in the old item ID embeddings to improve the ID embeddings of new items.

Inspired by ItemCF~\cite{linden2003amazon}, we consider that two items may be similar if they interact with a same user.
Moreover,~\cite{breese2013empirical} indicates that the more interactions a user has, the less he/she contributes to item similarity.
Hence, we design the item similarity score between item $i$ and item $j$ as following:
\begin{equation}
    Sim \left ( i,j \right ) = \frac{\sum_{a\in \left ( U\left ( i\right ) \bigcap U \left ( j\right )\right )}^{} \frac{1}{\log\left ( 1+\left | I\left ( a\right )\right |\right )}}{\sqrt{\left | U\left ( i\right )\right |\left | U\left ( j\right )\right |}},
\end{equation}
where $U\left ( i\right )$ is the set of the users who interacted with item $i$; $\left | U\left ( i\right )\right |$ is the number of users in $U\left ( i\right )$; $\left | I\left ( a\right )\right |$ is the number of the items which were interacted with by user $a$.
We keep the top-$K$ similar old items as set $T\left ( i\right )$ for item $i$, and the similar old items' weight can be expressed as:
\begin{equation}
    \alpha _{ij} = \frac{ Sim\left ( i,j\right )}{\sum_{k\in T\left ( i\right )}^{}  Sim\left ( i,k\right )}.
\end{equation}
Figure~\ref{fig:3} illustrates finding similar old items for a new item in B-EG.

Finally, B-EG generates the base embeddings for new items based on the ID embeddings of similar old items.
More specifically, for a new item $i$ we obtain its base embedding $\mathbf{v}_{BEG\left [ i \right ]}$ by computing the weighted sum of the ID embeddings of old items:
\begin{equation}
    \mathbf{v}_{BEG\left [ i \right ]} = \sum_{k\in T\left ( i\right )}^{} \alpha_{ik} \mathbf{v}_{k}.
\end{equation}

\subsection{S-EG: Shift Embedding Generator}
Existing models such as MetaEmb~\cite{pan2019warm} use item attribute features to generate meta embeddings for new item IDs.
However, these models do not pay attention to the collaborative information.
As with old items' ID embeddings, users' ID embeddings also contain useful collaborative information.
Therefore, in S-EG we generate a shift embedding for each new item by using its attribute features and the average interacted user ID embedding of the new item.
More specifically, as for new item $i$, we first compute its refined interacted users representation as:
\begin{equation}
    \mathbf{h}_{i}^{u} = \mathbf{W}_{u}Mean\left ( \left \{ \mathbf{u}_{j},j\in U\left ( i\right ) \right \}\right ),
\end{equation}
where $\mathbf{W}_{u}$ is a learnable parameter; $Mean\left ( \cdot  \right )$ is mean pooling.

Then, we compute its refined attribute features representation as:
\begin{equation}
    \mathbf{h}_{i}^{f} = \mathbf{W}_{f}\mathbf{Z}_{i},
\end{equation}
where $\mathbf{W}_{f}$ is a learnable parameter.

Finally, given its refined interacted users representation and attribute features representation, we obtain its shift embedding as:
\begin{equation}
    \mathbf{v}_{SEG\left [ i\right ]}=g_{\omega }\left ( \mathbf{h}_{i}^{u}, \mathbf{h}_{i}^{f} \right ),
\end{equation}
where $g_{\omega }\left ( \cdot  \right )$ is a MLP with parameters $\omega$.

\textbf{The training process of S-EG.}
Following the cold-start work~\cite{pan2019warm}, we view learning a shift embedding for each item as a task.
To achieve fast adaptation with limited interaction data,
we use a gradient-based meta learning method, which generalizes MAML~\cite{finn2017model}.
When training S-EG, we only update the parameters $\left \{ \mathbf{W}_{u},\mathbf{W}_{f},\omega  \right \}$ and freeze other parameters.
We use old items' interaction data to train S-EG.
Given an item $i$, we randomly select two disjoint minibatches of labeled data $\mathcal{D}^{a}_{i}$ and $\mathcal{D}^{b}_{i}$, each with $M$ samples.

We first make predictions on the first minibatch $\mathcal{D}^{a}_{i}$ as:
\begin{equation}
    \hat{y}_{aj} = f_{\theta }\left ( \mathbf{u}_{aj},\mathbf{Z}_{aj},\mathbf{v}_{BEG\left [ i\right ]} + 
 \mathbf{v}_{SEG\left [ i\right ]},\mathbf{Z}_{i} \right ),
\end{equation}
where the subscript $aj$ denotes the $j$-th sample from $\mathcal{D}^{a}_{i}$; $f_{\theta }\left ( \cdot \right )$ is the recommendation model.
Then, the average loss over these samples is calculated as following:
\begin{equation}
    loss_{a} = -\frac{1}{M}\sum_{j=1}^{M}y_{aj}\log\hat{y}_{aj}+\left ( 1-y_{aj} \right ) \log (1-\hat{y}_{aj}).
\end{equation}

We get a new embedding by computing the gradient of $loss_{a}$ \textit{w.r.t}. the initial shift embedding and taking a step of gradient descent:
\begin{equation}
    \mathbf{v}_{SEG\left [ i\right ]}^{'}=\mathbf{v}_{SEG\left [ i\right ]}-\eta \frac{\partial loss_{a}}{\partial \mathbf{v}_{SEG\left [ i\right ]}},
\end{equation}
where $\eta$ is the step size of gradient descent.

Next, we use the new embedding and make predictions on minibatch $\mathcal{D}^{b}_{i}$ as:
\begin{equation}
    \hat{y}_{bj} = f_{\theta }\left ( \mathbf{u}_{bj},\mathbf{Z}_{bj},\mathbf{v}_{BEG\left [ i\right ]} + 
 \mathbf{v}_{SEG\left [ i\right ]}^{'},\mathbf{Z}_{i} \right ).
\end{equation}
Then we calculate the average loss:
\begin{equation}
    loss_{b} = -\frac{1}{M}\sum_{j=1}^{M}y_{bj}\log\hat{y}_{bj}+\left ( 1-y_{bj} \right ) \log (1-\hat{y}_{bj}).
\end{equation}

The final loss function of the S-EG is:
\begin{equation}
    loss_{SEG} = \beta loss_{a} + \left ( 1 - \beta \right )loss_{b},
\end{equation}
where $\beta \in \left [ 0,1\right ]$ is a hyperparameter to balance the two losses.
The loss function considers two aspects:
1) The error of recommendation for the new items should be small.
2) The generated embeddings should adapt fast with limited interaction data.

\section{Experiments}

\subsection{Datasets}
We evaluate CoMeta on two public datasets MovieLens-1M and Taobao Ad.

\textbf{MovieLens-1M dataset\footnote{\url{http://www.grouplens.org/datasets/movielens/}}.}
It is one of the most well-known benchmark datasets, which contains 1 million movie rating records over thousands of movies and users.
Each movie can be seen as an item and has 4 features: ID, title, genre and year of release.
The associated features of each user include ID, age, gender and occupation.
We transform the ratings that are at least 4 to 1 and the others to 0.

\textbf{Taobao Ad dataset\footnote{\url{https://tianchi.aliyun.com/dataset/dataDetail?dataId=56}}.}
It is collected from the ad display logs in Taobao\footnote{\url{https://www.taobao.com/}} and contains 26 million click records from 1.14 million users in 8 days.
Each ad can be seen as an item, with features including its ad ID, category ID, campaign ID, brand ID and price.
Each user has 9 features: user ID, micro group ID, cms group ID, gender, age, consumption grade, shopping depth, occupation and city level.

\subsection{Backbones and Baselines}

\subsubsection{Backbones.}
The proposed CoMeta is model-agnostic and it can be applied upon various recommendation models.
Thus, we conduct experiments on the following backbones:

\begin{itemize}
    \item \textbf{Wide\&Deep}~\cite{cheng2016wide}.
    Wide\&Deep consists of logistic regression (LR) and DNNs which capture both low-order and high-order interactions.
    \item \textbf{DeepFM}~\cite{guo2017deepfm}.
    DeepFM combines factorization machine (FM) and DNNs.
    FM models first-order and second-order feature interactions, and DNNs model high-order feature interactions.
    \item \textbf{PNN}~\cite{qu2016product}.
    Product-Based Neural Network introduces a product layer into DNNs.
    We use two variants of PNN in the experiments:
    IPNN (PNN with an inner product layer) and OPNN (PNN with an outer product layer).
    \item \textbf{AFM}~\cite{xiao2017attentional}.
    Attentional Factorization Machine introduces the attention mechanism into FM, which can distinguish the importance of different feature interactions.
\end{itemize}

\subsubsection{Baselines.}
We choose some state-of-the-art (SOTA) meta-learning methods for the cold-start problem as baselines:
\begin{itemize}
    \item \textbf{MetaEmb}~\cite{pan2019warm}.
    Using a meta-learning approach, Meta-Embedding trains an embedding generator to generate good initial embeddings for new item IDs by using the attribute features of the new items.
    \item \textbf{GME}~\cite{ouyang2021learning}.
    Graph Meta Embedding generates desirable initial embeddings for new item IDs based on meta learning and graph neural networks.
    We use GME-A, which has the best performance among all GME models, as a baseline.
    \item \textbf{MWUF}~\cite{zhu2021learning}.
    Meta Warm Up Framework trains meta scaling and meta shifting networks to transform cold item ID embeddings into warm feature space.
    The scaling function transforms the cold ID embedding into the warm feature space, and the shifting function is able to produce more stable embeddings.
\end{itemize}

\begin{table}[htb]
\centering
\caption{Statistics of the datasets.}
\label{tab:1_v2}
\resizebox{\textwidth}{!}{%
\begin{tabular}{|l|lll|lll|}
\hline
\multirow{2}{*}{Dataset} & \multicolumn{3}{c|}{old items}                                                                                                                                                                                                                    & \multicolumn{3}{c|}{new items}                                                                                                                                                                         \\ \cline{2-7} 
                         & \multicolumn{1}{l|}{\# item IDs} & \multicolumn{1}{l|}{\begin{tabular}[c]{@{}l@{}}\# samples for pre-training\\  the recommendation model\end{tabular}} & \begin{tabular}[c]{@{}l@{}}\# samples for training\\  the cold-start model\end{tabular} & \multicolumn{1}{l|}{\# item IDs} & \multicolumn{1}{l|}{\begin{tabular}[c]{@{}l@{}}\# samples for\\  warm-up training\end{tabular}} & \begin{tabular}[c]{@{}l@{}}\# samples for\\  testing\end{tabular} \\ \hline
MovieLens-1M             & \multicolumn{1}{l|}{1,058}       & \multicolumn{1}{l|}{765,669}                                                                                         & 84,640                                                                                  & \multicolumn{1}{l|}{1,127}       & \multicolumn{1}{l|}{67,620}                                                                     & 123,787                                                           \\ \hline
Taobao Ad                & \multicolumn{1}{l|}{892}         & \multicolumn{1}{l|}{3,592,047}                                                                                       & 1,784,000                                                                               & \multicolumn{1}{l|}{540}         & \multicolumn{1}{l|}{810,000}                                                                    & 109,712                                                           \\ \hline
\end{tabular}%
}
\end{table}

\subsection{Experimental Settings}

\subsubsection{Dataset Splits.}
We group the items by their sizes following~\cite{pan2019warm} and~\cite{zhu2021learning}:
\begin{itemize}
    \item \textbf{Old items}:
    We regard the items whose number of samples is larger than a threshold $N_{old}$ as old items.
    For MovieLens-1M and Taobao Ad datasets, we use $N_{old}$ of 200 and 2000, respectively.
    \item \textbf{New items}:
    We regard the items whose number of samples is less than $N_{old}$ but larger than $N_{new}$ as new items.
    We sort the samples of each new item by timestamp and divide them into 4 groups. Subsequently, we use the first $3 \times K_{fold}$ samples as warm-a, warm-b and warm-c sets while the rest as test set.
    $N_{new}$ is set as $\{80, 1500\}$ and $K_{fold}$ is set as $\{20, 500\}$ for the two datasets, respectively.
\end{itemize}
Table \ref{tab:1_v2} shows the details of the datasets.

\subsubsection{Implementation Details.}
We set the embedding size for each feature as 16 and the balance hyperparameter $\beta$ as $0.1$.
The MLP in recommendation models uses the same structure with three hidden layers (64-64-64).
In addition, we set learning rate of 0.001 and optimize the parameters of the models with the Adam algorithm.
Following~\cite{pan2019warm}, the experiments are done with the following steps:
\begin{enumerate}[1)]
    \item Pre-train the base recommendation model with old item interaction data.
    \item Train extra modules (\textit{e.g.}, embedding generators, meta networks).
    \item Initialize the new item ID embeddings with the extra modules and calculate evaluation metrics on the test set.
    In this step, the new items do not have any interactions.
    \item Generate the better meta embeddings for new item IDs. In this step, the new items have a few interactions. \textbf{Note that only CoMeta needs this step.}
    \item Update the new item ID embeddings with warm-a set and calculate evaluation metrics on the test set.
    \item Update the new item ID embeddings with warm-b set and calculate evaluation metrics on the test set.
    \item Update the new item ID embeddings with warm-c set and calculate evaluation metrics on the test set.
\end{enumerate}
We denote the step 3 as cold phase, and the steps 5-7 as warm-a, warm-b and warm-c phases.
In the cold phase, as new items do not have any interactions, we can not calculate the base embeddings in B-EG and the average interacted user ID embeddings in S-EG.
To address this challenge, we take the average ID embeddings of global items and global users as the base embeddings and the average interacted user ID embeddings, respectively.
In the warm phase, unlike other methods that use the embeddings generated in the cold phase as the initial ID embeddings of new items, CoMeta regenerates the meta embeddings according to the interaction data in the warm phase.
Following~\cite{pan2019warm}, for each training step, we only update the model parameters for 1 epoch.

\subsubsection{Evaluation Metrics.}
We adopt two widely used evaluation metrics to evaluate the performance:
1) \textbf{AUC.}
Area under the ROC curve.
The larger the better.
2) \textbf{Logloss.}
Binary cross-entropy loss.
The smaller the better.

\begin{table*}[htb]
\centering
\caption{Experimental results on two public datasets.
The best result is bolded.}
\label{tab:2}
\resizebox{\textwidth}{!}{%
\begin{tabular}{|c|c|c||cc||cc||cc||cc|}
\hline
\multirow{2}{*}{Dataset}       & \multirow{2}{*}{Backbone} & \multirow{2}{*}{Method} & \multicolumn{2}{c||}{Cold phase}   & \multicolumn{2}{c||}{Warm-a phase} & \multicolumn{2}{c||}{Warm-b phase} & \multicolumn{2}{c|}{Warm-c phase} \\
                               &                             &                         & AUC             & Logloss         & AUC             & Logloss         & AUC             & Logloss         & AUC             & Logloss         \\ \hline
\multirow{10}{*}{MovieLens-1M} & \multirow{5}{*}{DeepFM}     & DeepFM                  & 0.7043          & 0.6660          & 0.7311          & 0.6370          & 0.7520          & 0.6126          & 0.7679          & 0.5920          \\
                               &                             & MetaEmb                 & 0.7120          & 0.6508          & 0.7380          & 0.6235          & 0.7577          & 0.6011          & 0.7724          & 0.5826          \\
                               &                             & GME                     & \textbf{0.7168} & 0.6443          & 0.7405          & 0.6188          & 0.7588          & 0.5978          & 0.7727          & 0.5804          \\
                               &                             & MWUF                    & 0.7099          & 0.6598          & 0.7471          & 0.6220          & 0.7686          & 0.5918          & 0.7821          & 0.5702          \\ \cdashline{3-11}
                               &                             & CoMeta                 & 0.7106          & \textbf{0.6381} & \textbf{0.7789} & \textbf{0.5706} & \textbf{0.7861} & \textbf{0.5607} & \textbf{0.7917} & \textbf{0.5526} \\ \cline{2-11} 
                               & \multirow{5}{*}{Wide\&Deep} & Wide\&Deep              & 0.7026          & 0.6686          & 0.7308          & 0.6386          & 0.7522          & 0.6139          & 0.7678          & 0.5936          \\
                               &                             & MetaEmb                 & 0.7108          & 0.6524          & 0.7376          & 0.6247          & 0.7574          & 0.6025          & 0.7718          & 0.5845          \\
                               &                             & GME                     & \textbf{0.7158} & 0.6435          & 0.7397          & 0.6182          & 0.7580          & 0.5978          & 0.7716          & 0.5812          \\
                               &                             & MWUF                    & 0.7067          & 0.6636          & 0.7441          & 0.6254          & 0.7664          & 0.5947          & 0.7797          & 0.5735          \\ \cdashline{3-11}
                               &                             & CoMeta                 & 0.7081          & \textbf{0.6375} & \textbf{0.7789} & \textbf{0.5695} & \textbf{0.7854} & \textbf{0.5609} & \textbf{0.7903} & \textbf{0.5538} \\ \hline
\multirow{10}{*}{Taobao Ad}    & \multirow{5}{*}{DeepFM}     & DeepFM                  & 0.5740          & 0.2270          & 0.5910          & 0.2244          & 0.6025          & 0.2229          & 0.6124          & 0.2218          \\
                               &                             & MetaEmb                 & 0.5770          & 0.2263          & 0.5934          & 0.2239          & 0.6042          & 0.2226          & 0.6138          & 0.2216          \\
                               &                             & GME                     & \textbf{0.5787} & 0.2260          & 0.5935          & 0.2238          & 0.6034          & 0.2226          & 0.6125          & 0.2216          \\
                               &                             & MWUF                    & 0.5779          & 0.2259          & 0.5932          & 0.2247          & 0.6020          & 0.2234          & 0.6107          & 0.2223          \\ \cdashline{3-11}
                               &                             & CoMeta                 & 0.5781          & \textbf{0.2258} & \textbf{0.5982} & \textbf{0.2234} & \textbf{0.6078} & \textbf{0.2223} & \textbf{0.6165} & \textbf{0.2213} \\ \cline{2-11} 
                               & \multirow{5}{*}{Wide\&Deep} & Wide\&Deep              & 0.5952          & 0.2257          & 0.6098          & 0.2234          & 0.6195          & 0.2220          & 0.6281          & 0.2208          \\
                               &                             & MetaEmb                 & 0.5970          & 0.2246          & 0.6111          & 0.2227          & 0.6204          & 0.2215          & 0.6287          & 0.2205          \\
                               &                             & GME                     & \textbf{0.6001} & 0.2240          & 0.6134          & 0.2223          & 0.6220          & 0.2212          & 0.6297          & 0.2203          \\
                               &                             & MWUF                    & 0.5974          & 0.2236          & 0.6102          & 0.2223          & 0.6188          & 0.2213          & 0.6271          & 0.2204          \\ \cdashline{3-11} 
                               &                             & CoMeta                 & 0.5974          & \textbf{0.2235} & \textbf{0.6153} & \textbf{0.2219} & \textbf{0.6231} & \textbf{0.2210} & \textbf{0.6304} & \textbf{0.2201} \\ \hline
\end{tabular}%
}
\end{table*}

\subsection{Experimental Results}

\subsubsection{Overall Performance.}
We conduct experiments on two public datasets and choose Wide\&Deep, DeepFM as backbones.
Meanwhile, we compare CoMeta with three SOTA cold-start methods: MetaEmb, GME, MWUF.
For each method, we run it five times and report the average result in Table \ref{tab:2}.

\textbf{The cold-start phase.}
MetaEmb and GME have better performance than the backbones, which indicates the attribute features can contribute useful information to item IDs.
In addition, GME performs much better than MetaEmb because the neighbor items contain more information.
MWUF takes the average embeddings of global items to initialize the new item ID embeddings, which is also helpful.
Although we focus on the warm-up phase, the proposed CoMeta can improve the new item ID embeddings in the cold-start phase as well.

\textbf{The warm-up phase.}
Since MetaEmb and GME both learn desirable initial embeddings for new item IDs in the cold-start phase, they still perform better than the backbones in the warm-up phase.
MWUF performs well on MovieLens-1M, but has poor prediction on Taobao Ad.
We guess the reason is that the attribute features of new items are not well-trained on Taobao Ad, which leads to MWUF can not correctly transform cold item ID embeddings into warm feature space.
It is observed that our CoMeta achieves the best performance among all methods on the two datasets.
Because we generate better meta embeddings, which contain much useful collaborative information.

\begin{table*}[htb]
\centering
\caption{The results of ablation study.}
\label{tab:3}
\resizebox{\textwidth}{!}{%
\begin{tabular}{|c|c||cc||cc||cc||cc|}
\hline
\multirow{2}{*}{Dataset}      & \multirow{2}{*}{Method} & \multicolumn{2}{c||}{Cold phase}   & \multicolumn{2}{c||}{Warm-a phase} & \multicolumn{2}{c||}{Warm-b phase} & \multicolumn{2}{c|}{Warm-c phase} \\
                              &                         & AUC             & Logloss         & AUC             & Logloss         & AUC             & Logloss         & AUC             & Logloss         \\ \hline
\multirow{4}{*}{MovieLens-1M} & DeepFM                  & 0.7043          & 0.6660          & 0.7311          & 0.6370          & 0.7520          & 0.6126          & 0.7679          & 0.5920          \\
                              & CoMeta (without B-EG)   & \textbf{0.7118} & 0.6475          & 0.7378          & 0.6227          & 0.7576          & 0.6007          & 0.7723          & 0.5824          \\
                              & CoMeta (without S-EG)   & 0.7045          & 0.6636          & 0.7743          & 0.5884          & 0.7823          & 0.5747          & 0.7887          & 0.5631          \\
                              & CoMeta                 & 0.7106          & \textbf{0.6381} & \textbf{0.7789} & \textbf{0.5706} & \textbf{0.7861} & \textbf{0.5607} & \textbf{0.7917} & \textbf{0.5526} \\ \hline
\multirow{4}{*}{Taobao Ad}    & DeepFM                  & 0.5740          & 0.2270          & 0.5910          & 0.2244          & 0.6025          & 0.2229          & 0.6124          & 0.2218          \\
                              & CoMeta (without B-EG)   & \textbf{0.5781} & 0.2261          & 0.5945          & 0.2238          & 0.6051          & 0.2225          & 0.6145          & 0.2215          \\
                              & CoMeta (without S-EG)   & 0.5779          & 0.2259          & 0.5948          & 0.2239          & 0.6049          & 0.2226          & 0.6140          & 0.2216          \\
                              & CoMeta                 & \textbf{0.5781} & \textbf{0.2258} & \textbf{0.5982} & \textbf{0.2234} & \textbf{0.6078} & \textbf{0.2223} & \textbf{0.6165} & \textbf{0.2213} \\ \hline
\end{tabular}%
}
\end{table*}

\subsubsection{Ablation Study.}
To explore the impact of different components in CoMeta, we conduct an ablation study on the two datasets with DeepFM:
1) DeepFM: the backbone.
2) CoMeta (without B-EG): CoMeta without the B-EG.
3) CoMeta (without S-EG): CoMeta without the S-EG.
4) CoMeta: the overall framework.
As shown in Table \ref{tab:3}, an interesting finding is that the B-EG contributes more than the S-EG on MovieLens-1M, but they contribute similarly on Taobao Ad.
Generally, for a new item, its old similar items contain more useful information than its attribute features and interacted users.
But Taobao Ad has more noisy interactions than MovieLens-1M, and we can not exactly calculate the base embeddings for new items on Taobao Ad.
The overall CoMeta achieves the best performance, which shows each component is important for CoMeta.

\begin{figure}
    \centering
	\includegraphics[width=1\linewidth]{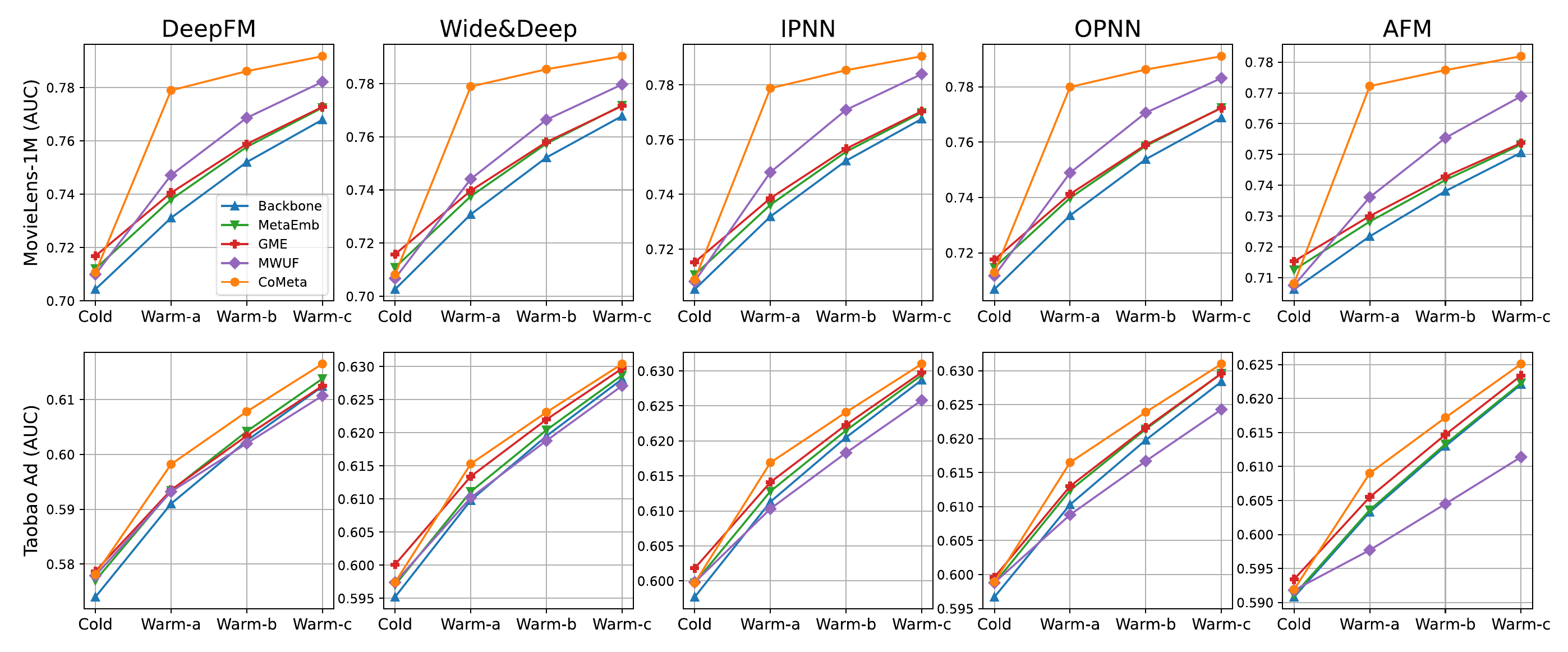} 
	\caption{AUC on two datasets, over five popular backbones.}
	\label{fig:4} 
\end{figure}

\subsubsection{Method Compatibility.}
Since CoMeta is model-agnostic, we conduct experiments upon more backbones.
Figure \ref{fig:4} shows the results on five representative backbones (DeepFM, Wide\&Deep, IPNN, OPNN, AFM) and two datasets.
We observe that our CoMeta can significantly improve the recommendation performance on new items with different backbones.
In general, the performance of AFM is worse than other backbones, because AFM can not capture high-order interactions.

\section{Conclusion}
In this paper, we focus on addressing the item cold-start problem.
We propose CoMeta that generates desirable meta embeddings with collaborative information for new item IDs.
The experimental results on two datasets demonstrate that our proposed CoMeta outperforms SOTA cold-start methods and has good compatibility.
In the future, we would like to extend our work to more scenarios (\textit{e.g.}, user cold-start problem).

%
%
%
%

\bibliographystyle{splncs04}
\bibliography{samplepaper}

\begin{thebibliography}{10}
\providecommand{\url}[1]{\texttt{#1}}
\providecommand{\urlprefix}{URL }
\providecommand{\doi}[1]{https://doi.org/#1}

\bibitem{breese2013empirical}
Breese, J.S., Heckerman, D., Kadie, C.: Empirical analysis of predictive
  algorithms for collaborative filtering. arXiv preprint arXiv:1301.7363
  (2013)

\bibitem{cheng2016wide}
Cheng, H.T., Koc, L., Harmsen, J., Shaked, T., Chandra, T., Aradhye, H.,
  Anderson, G., Corrado, G., Chai, W., Ispir, M., et~al.: Wide \& deep learning
  for recommender systems. In: Proceedings of the 1st workshop on deep learning
  for recommender systems. pp. 7--10 (2016)

\bibitem{dai2021poso}
Dai, S., Lin, H., Zhao, Z., Lin, J., Wu, H., Wang, Z., Yang, S., Liu, J.: Poso:
  Personalized cold start modules for large-scale recommender systems. arXiv
  preprint arXiv:2108.04690  (2021)

\bibitem{du2020learn}
Du, X., Wang, X., He, X., Li, Z., Tang, J., Chua, T.S.: How to learn item
  representation for cold-start multimedia recommendation? In: Proceedings of
  the 28th ACM International Conference on Multimedia. pp. 3469--3477 (2020)

\bibitem{finn2017model}
Finn, C., Abbeel, P., Levine, S.: Model-agnostic meta-learning for fast
  adaptation of deep networks. In: International conference on machine
  learning. pp. 1126--1135. PMLR (2017)

\bibitem{guo2017deepfm}
Guo, H., Tang, R., Ye, Y., Li, Z., He, X.: Deepfm: a factorization-machine
  based neural network for ctr prediction. arXiv preprint arXiv:1703.04247
  (2017)

\bibitem{kingma2013auto}
Kingma, D.P., Welling, M.: Auto-encoding variational bayes. arXiv preprint
  arXiv:1312.6114  (2013)

\bibitem{lee2019melu}
Lee, H., Im, J., Jang, S., Cho, H., Chung, S.: Melu: Meta-learned user
  preference estimator for cold-start recommendation. In: Proceedings of the
  25th ACM SIGKDD International Conference on Knowledge Discovery \& Data
  Mining. pp. 1073--1082 (2019)

\bibitem{linden2003amazon}
Linden, G., Smith, B., York, J.: Amazon. com recommendations: Item-to-item
  collaborative filtering. IEEE Internet computing  \textbf{7}(1),  76--80
  (2003)

\bibitem{ouyang2021learning}
Ouyang, W., Zhang, X., Ren, S., Li, L., Zhang, K., Luo, J., Liu, Z., Du, Y.:
  Learning graph meta embeddings for cold-start ads in click-through rate
  prediction. In: Proceedings of the 44th International ACM SIGIR Conference on
  Research and Development in Information Retrieval. pp. 1157--1166 (2021)

\bibitem{pan2019warm}
Pan, F., Li, S., Ao, X., Tang, P., He, Q.: Warm up cold-start advertisements:
  Improving ctr predictions via learning to learn id embeddings. In:
  Proceedings of the 42nd International ACM SIGIR Conference on Research and
  Development in Information Retrieval. pp. 695--704 (2019)

\bibitem{qu2016product}
Qu, Y., Cai, H., Ren, K., Zhang, W., Yu, Y., Wen, Y., Wang, J.: Product-based
  neural networks for user response prediction. In: 2016 IEEE 16th
  International Conference on Data Mining (ICDM). pp. 1149--1154. IEEE (2016)

\bibitem{rendle2010factorization}
Rendle, S.: Factorization machines. In: 2010 IEEE International conference on
  data mining. pp. 995--1000. IEEE (2010)

\bibitem{rendle2012bpr}
Rendle, S., Freudenthaler, C., Gantner, Z., Schmidt-Thieme, L.: Bpr: Bayesian
  personalized ranking from implicit feedback. arXiv preprint arXiv:1205.2618
  (2012)

\bibitem{DBLP:conf/ijcai/RongHC22}
Rong, D., He, Q., Chen, J.: Poisoning deep learning based recommender model in
  federated learning scenarios. In: Raedt, L.D. (ed.) Proceedings of the
  Thirty-First International Joint Conference on Artificial Intelligence,
  {IJCAI} 2022, Vienna, Austria, 23-29 July 2022. pp. 2204--2210. ijcai.org
  (2022). \doi{10.24963/ijcai.2022/306},
  \url{https://doi.org/10.24963/ijcai.2022/306}

\bibitem{rong2022fedrecattack}
Rong, D., Ye, S., Zhao, R., Yuen, H.N., Chen, J., He, Q.: Fedrecattack: Model
  poisoning attack to federated recommendation. In: 2022 IEEE 38th
  International Conference on Data Engineering (ICDE). pp. 2643--2655. IEEE
  (2022)

\bibitem{schein2002methods}
Schein, A.I., Popescul, A., Ungar, L.H., Pennock, D.M.: Methods and metrics for
  cold-start recommendations. In: Proceedings of the 25th annual international
  ACM SIGIR conference on Research and development in information retrieval.
  pp. 253--260 (2002)

\bibitem{shi2019adaptive}
Shi, S., Zhang, M., Yu, X., Zhang, Y., Hao, B., Liu, Y., Ma, S.: Adaptive
  feature sampling for recommendation with missing content feature values. In:
  Proceedings of the 28th ACM International Conference on Information and
  Knowledge Management. pp. 1451--1460 (2019)

\bibitem{volkovs2017dropoutnet}
Volkovs, M., Yu, G., Poutanen, T.: Dropoutnet: Addressing cold start in
  recommender systems. Advances in neural information processing systems
  \textbf{30} (2017)

\bibitem{wang2021preference}
Wang, L., Jin, B., Huang, Z., Zhao, H., Lian, D., Liu, Q., Chen, E.:
  Preference-adaptive meta-learning for cold-start recommendation. In: IJCAI.
  pp. 1607--1614 (2021)

\bibitem{wang2017deep}
Wang, R., Fu, B., Fu, G., Wang, M.: Deep \& cross network for ad click
  predictions. In: Proceedings of the ADKDD'17, pp.~1--7 (2017)

\bibitem{xiao2017attentional}
Xiao, J., Ye, H., He, X., Zhang, H., Wu, F., Chua, T.S.: Attentional
  factorization machines: Learning the weight of feature interactions via
  attention networks. arXiv preprint arXiv:1708.04617  (2017)

\bibitem{zhang2021unbert}
Zhang, Q., Li, J., Jia, Q., Wang, C., Zhu, J., Wang, Z., He, X.: Unbert:
  User-news matching bert for news recommendation. In: IJCAI. pp. 3356--3362
  (2021)

\bibitem{zhao2022joint}
Zhao, K., Zheng, Y., Zhuang, T., Li, X., Zeng, X.: Joint learning of e-commerce
  search and recommendation with a unified graph neural network. In:
  Proceedings of the Fifteenth ACM International Conference on Web Search and
  Data Mining. pp. 1461--1469 (2022)

\bibitem{zhao2022improving}
Zhao, X., Ren, Y., Du, Y., Zhang, S., Wang, N.: Improving item cold-start
  recommendation via model-agnostic conditional variational autoencoder. arXiv
  preprint arXiv:2205.13795  (2022)

\bibitem{zhou2019deep}
Zhou, G., Mou, N., Fan, Y., Pi, Q., Bian, W., Zhou, C., Zhu, X., Gai, K.: Deep
  interest evolution network for click-through rate prediction. In: Proceedings
  of the AAAI conference on artificial intelligence. vol.~33, pp. 5941--5948
  (2019)

\bibitem{zhou2018deep}
Zhou, G., Zhu, X., Song, C., Fan, Y., Zhu, H., Ma, X., Yan, Y., Jin, J., Li,
  H., Gai, K.: Deep interest network for click-through rate prediction. In:
  Proceedings of the 24th ACM SIGKDD international conference on knowledge
  discovery \& data mining. pp. 1059--1068 (2018)

\bibitem{zhu2021learning}
Zhu, Y., Xie, R., Zhuang, F., Ge, K., Sun, Y., Zhang, X., Lin, L., Cao, J.:
  Learning to warm up cold item embeddings for cold-start recommendation with
  meta scaling and shifting networks. In: Proceedings of the 44th International
  ACM SIGIR Conference on Research and Development in Information Retrieval.
  pp. 1167--1176 (2021)

\end{thebibliography}


\begin{thebibliography}{8}
\bibitem{ref_article1}
Author, F.: Article title. Journal \textbf{2}(5), 99--110 (2016)

\bibitem{ref_lncs1}
Author, F., Author, S.: Title of a proceedings paper. In: Editor,
F., Editor, S. (eds.) CONFERENCE 2016, LNCS, vol. 9999, pp. 1--13.
Springer, Heidelberg (2016). \doi{10.10007/1234567890}

\bibitem{ref_book1}
Author, F., Author, S., Author, T.: Book title. 2nd edn. Publisher,
Location (1999)

\bibitem{ref_proc1}
Author, A.-B.: Contribution title. In: 9th International Proceedings
on Proceedings, pp. 1--2. Publisher, Location (2010)

\bibitem{ref_url1}
LNCS Homepage, \url{http://www.springer.com/lncs}. Last accessed 4
Oct 2017
\end{thebibliography}

\end{document}